\documentclass{iaus}
\usepackage{graphicx}

\title[JD 11.~~Spectropolarimetry of Rigel] 
{Searching for Weak or Complex Magnetic \\ Fields in Polarized Spectra of Rigel}

\author[M. Shultz, G. A. Wade, \& the MiMeS Collaboration]   
{M. Shultz$^1$ $^2$, G. A. Wade$^2$, C. Neiner$^3$, N. Manset$^4$, V. Petit$^2$ $^5$, J. Grunhut$^1$ $^2$, E. Guinan$^6$, D. Hanes$^1$
 \and the MiMeS Collaboration}

\affiliation{$^1$Queen's University, Canada $^2$Royal Military College, Canada $^3$Paris-Meudon Observatory \\ $^4$ Canada-France-Hawaii Telescope Corporation $^5$West Chester U $^6$Villanova U}

\pubyear{2010}
\volume{272}  
\pagerange{119--126}
\jname{Active OB Stars: structure, evolution, mass loss and critical limits}
\editors{A.C. Editor, B.D. Editor \& C.E. Editor, eds.}
\begin{document}

\maketitle

\begin{abstract}
Seventy-eight high-resolution Stokes V, Q and U spectra of the B8Iae supergiant Rigel were obtained with the ESPaDOnS spectropolarimeter at CFHT and its clone NARVAL at TBL in the context of the Magnetism in Massive Stars (MiMeS) Large Program, in order to scrutinize this core-collapse supernova progenitor for evidence of weak and/or complex magnetic fields. In this paper we describe the reduction and analysis of the data, the constraints obtained on any photospheric magnetic field, and the variability of photospheric and wind lines.
\end{abstract}

\firstsection 
\section{Introduction: Physical Parameters and Observations}

Rigel: a blue supergiant, the closest and most readily studied Type II supernova progenitor, and a known $\alpha$ Cygni variable, is the subject of a global monitoring campaign known as the 'Rigel-thon', involving long-term spectroscopic monitoring, Microvariability and Oscillations in STars (MOST) space photometry, and spectropolarimetry.

Like most OB stars, $\beta$ Ori A shows no sign of an easily detected magnetic field, however apparent brightness and sharp spectral lines make it practical to ask if the star possesses a weak or complex field geometry which might be revealed within a high resolution data set. Thus, over the epoch 09/2009--02/2010, 65 Stokes V (circular polarization) and 13 Stokes Q and U (linear polarization) spectra spanning 370-1000 nm with a mean resolving power R $\sim$ 65000 at 500 nm, were taken with the ESPaDOnS spectropolarimeter at CFHT and its clone Narval at TBL. Integration times were typically of a few seconds duration. The densest spectropolarimetric sampling was concurrent with the collection of MOST data.

The physical radius was determined from the interferometric angular diameter, $\theta_D$ = 2.76 $\pm$ 0.01 mas (\cite[Aufdenberg \textit{et al}, 2008]{Aufdenberg_2008}) together with a distance of 240 $\pm$ 50 pc calculated using the Hipparcos parallax of 4.22 $\pm$ 0.81 mas: thus R = 70 $\pm$ 14 R$_\odot$. The star appears to be a slow rotator, with \textit{v}sin\textit{i} = 36 $\pm$ 5 km/s (\cite[Przybilla \textit{et al}. 2009]{Przybilla_2009}), giving an upper bound on the rotation period P$_{rot}$ of $\sim$ 98 d; calculation of the breakup velocity ($\sim$ 250 km/s) provides a lower limit of $\sim$ 14 d. 

\section{Results and Analysis}

Least Squares Deconvolution (LSD) was employed to extract high S/N ratio mean Stokes I, V, and diagnostic N profiles from the circular polarization spectra.  The LSD line mask was cleaned to eliminate contamination from telluric, emission and Balmer lines, and to remove weak or apparently absent lines. Ultimately $\sim$ 90 lines remained, and their  weights were empirically adjusted to reflect observed line depths. The typical S/N ratio in Stokes V mean profiles was $\sim$ 20,000. No significant signal was detected in either Stokes V or diagnostic N. 

Each LSD profile was then analyzed to determine the longitudinal magnetic field B$_l$. No significant longitudinal field was detected, with a median 1-$\sigma$ uncertainty in individual measurements of 13 G. The distribution of B$_l$ values inferred from Stokes V is statistically identical to that inferred from diagnostic N. The measured B$_l$ was then compared to a grid of synthetic longitudinal field curves corresponding to dipoles with 0$^{\circ}$ $\leq \beta \leq$ 90$^{\circ}$, 0$^{\circ}$ $\leq i \leq$ 90$^{\circ}$ (with the data folded according to the theoretical maximum P$_{rot}$ = 93 d at i = 90$^{\circ}$ and progressively shorter periods at smalled i, with ten different phase offsets tested at each period), and polar field strengths B$_d$ from 0 to $\sim$ 3 kG. For (i = 90$^{\circ}$, $\beta$ = 90$^{\circ}$) the maximum B$_d$ compatible with the data at 3-$\sigma$ confidence is $\sim$ 20 G, while B$_d$ is constrained below $\sim$ 50 G for intermediate values of i and $\beta$. Fields at this level, if present within the photosphere, remain capable of strongly influencing the wind (\cite[ud-Doula \& Owocki, 2002]{udDoula_Owocki_2002}), with a wind magnetic confinement parameter $\eta_* \sim$  2 -- 90, assuming $\dot M$ $\sim$ 10$^{-7}$--10$^{-6}$ M$_\odot$/y (\cite[Barlow \& Cohen 1977]{Barlow_Cohen_1977}, \cite[Abbot 1980]{Abbot_1980}, \cite[Puls 2008]{Puls_2008}) and \textit{v}$_\infty \sim$ 400--600 km/s (\cite[Bates, 1980]{Bates_1980}). 

Rigel is a long-known $\alpha$ Cygni variable (\cite[Sanford, 1947]{Sanford_1947}), with significant line profile variability (LPV) in H$\alpha$ as well as various metal lines, which may be associated with any or all of: mass loss events, photospheric spots, corotating interacting structures, and/or g- or p-mode pulsations (\cite[Kaufer, 1996a]{Kaufer_1996a}, \cite[1997]{Kaufer_1997}). Distinct LPV is seen in H$\alpha$ as compared to metal lines: H$\alpha$ is in strong emission, variable over a broad velocity range and apparently aperiodic (consistent with earlier spectroscopic monitoring (\cite[Kaufer 1996a]{Kaufer_1996a}, \cite[1996b]{Kaufer_1996b}, \cite[Israelian 1997]{Israelian_1997})); metal lines showed little apparent emission excess, but their variability was  suggestive of periodic behaviour. Amongst the most complexly variable of the metal lines is the O triplet at 777 nm.

\section{Conclusions \& Future Work}

No evidence of a magnetic field is obtained in 65 high precision Stokes V observations of Rigel. Significant variability is observed in numerous spectral lines, with some suggestion of periodicity on the order of $\sim$ 1 month in metallic lines. Further modeling of Stokes V profiles must be performed to obtain quantitative constraints on various potential field topologies, e.g. the dynamo-generated field proposed by Cantiello (\cite[2009]{Cantiello_2009}); a more rigorous analysis of LPV may help to identify periodic behaviour.

\end{document}